\documentclass[a4paper]{jpconf}
\usepackage{graphicx}
\begin{document}
\title{Diffusive shock acceleration in relativistic, oblique shocks}

\author{Allard Jan van Marle$^1$, Artem Bohdan$^{2,3}$, Anabella Araudo$^{1,4}$, Fabien Casse$^5$, Alexandre Marcowith$^4$}

\address{$^1$ Extreme Light Infrastructure ERIC, ELI Beamlines Facility, Za Radnici 835, CZ-25241 Dolni Brezany, Czech Republic }

\address{$^2$ Max-Planck-Institut für Plasmaphysik, Boltzmannstr. 2, DE-85748 Garching, Germany}

\address{$^3$ Excellence Cluster ORIGINS, Boltzmannstr. 2, DE-85748 Garching, Germany}

\address{$^4$ Laboratoire Univers et Particules de Montpellier (LUPM) Universit\'e Montpellier, CNRS/IN2P3, CC72, Place Eug\'ene Bataillon, F-34095 Montpellier Cedex5, France}

\address{$^5$ Universit\'e de Paris, AstroParticle \& Cosmologie, CNRS CEA, Observatoire de Paris, PSL Research University, CNES, F-75013 Paris, France}

\ead{ajvanmarle@gmail.com}

\begin{abstract}
Cosmic rays are charged particles that are accelerated to relativistic speeds by astrophysical shocks. Numerical models have been
successful in confirming the acceleration process for (quasi-)parallel shocks, which have the magnetic field aligned with the
direction of the shock motion. However, the process is less clear when it comes to (quasi-)perpendicular shocks, where the field makes a large angle with the shock-normal. For such shocks, the angle between the magnetic field and flow ensures that
only highly energetic particles can travel upstream at all, reducing the upstream current. This process is further inhibited for relativistic shocks, since the shock can become superluminal when the required particle velocity exceeds the speed of light, effectively inhibiting any upstream particle flow.
In order to determine whether such shocks can accelerate particles, we use the particle-in-cell (PIC) method to determine what fraction of particles gets reflected initially at the
shock. We then use this as input for a new simulation that combines the PIC method with grid-based magnetohydrodynamics to
follow the acceleration (if any) of the particles over a larger time-period in a two-dimensional grid. We find that quasi-perpendicular, relativistic shocks are capable of accelerating particles through the DSA process, provided that the shock has a sufficiently high Alfv{\'e}nic Mach number.
\end{abstract}

\section{Introduction}
Cosmic rays are charged particles that have been accelerated to relativistic speeds. It is generally accepted that the acceleration process occurs at astrophysical shocks, with the particles gaining energy through repeated shock crossings. One of the main processes involved is that of Diffusive Shock Acceleration (DSA) [1], also known as Fermi~1 acceleration. This involves the deflection of charged particles by the upstream and downstream magnetic field, which requires magnetic instabilities. 
One of the sources of such instabilities is the presence of the charged particles themselves. 
The current generated by these particles as they move relative to the thermal gas can trigger the "streaming instability", which twists the magnetic field [2,3]. Because the wavelength of this instability scales with the current, it can continue to interact with the particles as they are accelerated, making the process both self-induced and self-sustaining. 

\subsection{Diffusive Shock Acceleration in oblique shocks}
The interaction between charged, non-thermal particles and a magnetic field and the DSA that follows has been studied in considerable detail in the case of a non-relativistic parallel shock, where the magnetic field is aligned with the shock-normal [4,5,6] and the motion of the gas does not approach the speed of light. However, the situation is less clear in the case of oblique, relativistic shocks. [4,5] found that at an angle of more than approximately 60$^o$, the upstream current became too weak to trigger instabilities and therefore DSA could no longer occur. By contrast, [6] did find DSA. However, these simulations assumed that both parallel and oblique shocks would inject non-thermal particles into the upstream medium at the same rate. 

In [7], the problem was further investigated by combining both particle-in-cell (PIC) and PIC-MHD (a combination of PIC and grid-based magnetohydrodynamics). Using a PIC code to model the shock itself, combined with PIC-MHD to model large-scale, long-term behaviour, it was determined that the occurrence of streaming instability, and the DSA that it causes, depends on the Alfv{\'e}nic Mach number of the shock. For non-thermal particles to trigger an instability, their energy density has to exceed the energy density of the magnetic field. In the upstream medium, this is a problem. To travel upstream, the particle has to exceed the velocity at which the upstream medium advects toward the shock.
Particles follow the magnetic field lines, which, for an oblique shock, means that the particle velocity has to exceed $v_s /\tan{\theta_B}$, where $v_s$ is the velocity of the upstream medium in the shock rest-frame; and $\theta_B$ is the angle between the magnetic field and the shock normal. The larger $\theta_B$, the higher the required velocity, which means that only the fastest particles  -a small fraction of the non-thermal particles that are generated at the shock-will be able to travel upstream. 
As a result, they can only perturb the magnetic field if that field is sufficiently weak, which translates to a high Alfv{\'e}nic Mach number. [7] showed that for $\theta_B\,=\,60^o$ the upstream Alfv{\'e}nic Mach number has to be at least $\simeq\,30$ for even partial DSA to take place, and only at Alfv{\'e}nic Mach numbers exceeding 100 does the process become efficient. Furthermore, this dependence is extremely strong: for $\theta_B\,=\,70^o$, the minimum Alfv{\'e}nic Mach number would be around 1000.

\subsection{Can DSA occur in oblique, relativistic shocks?}
The model conclusions drawn in the  previous section only apply to non-relativistic shocks. 
If the shock speed approaches the speed of light, a second obstacle to DSA appears: While particles have to move at a speed larger than $v_s /\tan{\theta_B}$ to travel into the upstream medium, at the same time, they still need to obey $v\,<\,c$, with $c$ the light velocity. So, for any shock where $v_s /\tan{\theta_B}\geq c$, (also referred to as super-luminal shocks) no particle will be able to move upstream, thereby precluding the possibility of DSA. 

Therefore, DSA in oblique, relativistic shocks can only occur within a relatively narrow parameter space, where the shock's Mach number is high enough that upstream particles can trigger instabilities, and the obliquity is not so large that the shock becomes super-luminal. Here it should also be noted that at relativistic speeds, the Lorentz contraction causes the obliquity in the shock rest-frame to be larger than in the co-moving frame, and it is the obliquity in the shock rest-frame that determines whether the shock is super-luminal. 
We will now show the preliminary results of our simulations of such shocks. 

\section{PIC simulations}
For our simulations, we repeat the process described in [7], which involves two sets of simulations. In the first, PIC is used to model the structure of the shock and determine the injection rate of non-thermal particles at the shock front. This data is then used in PIC-MHD models to determine the evolution of the plasma and its magnetic field in a long-term, 2-D simulation.

The PIC simulations are performed with a fully relativistic electromagnetic 2-D PIC code based on the TRISTAN code [8,9].
Shocks are initialized using the reflecting-wall setup, where a beam of plasma is aimed at a reflective boundary, and the resulting shock is traced as it moves back into the upstream medium. 
For these simulations, we choose to use an out-of-plane magnetic field configuration, so the magnetic field lies in the x-z, plane, whereas the simulation box is in the x-y plane. We use an upstream Lorentz-factor $\Gamma\,=\,1.86$ and vary $\theta_B$ from 0 to 33 in the upstream rest-frame, which puts it at 0 to 60 degrees in the shock rest-frame. This covers the full parameter space from parallel to superluminal shocks The Alfv{\'e}nic and sonic Mach numbers are 20 and 22 respectively to give us a plasma-$\beta$\,=\,1. 
The results of these simulations were presented in [10], and showed the dependence of the particle injection rate in the upstream medium as a function of the shock obliquity, starting high for e parallel shock, but decreasing with an increased obliquity. For the superluminal shock, no particles were injected into the upstream medium.

\section{PIC-MHD simulations}
From the injection rate obtained in the PIC model, we start our PIC-MHD model. The PIC-MHD method, which was described in [11], assumes that a plasma can be described as a combination of two components: a thermal plasma, which can be treated as a fluid through traditional grid-based magnetohydrodynamics (MHD) and a non-thermal component that is treated as a collection of particles, modelled through PIC. These two components exist in the same time and space and interact with each other through the Lorentz force (which describes the force, exerted by the electro-magnetic field and the particles on each other), and a modified version of Ohm's law that takes into account an additional component of the electric field, generated by the non-thermal current. 
This method was implemented in the existing MPI-AMRVAC code [12], as described in [6]. Although this PIC-MHD code was initially limited to non-relativistic MHD, it has since been extended to include relativistic magneto-hydrodynamics. 
By combining MHD and PIC, we can model the interaction between non-thermal particles and an electromagnetic field at a lower computational cost than if we used only PIC. This allows us to use a larger simulation box and run the simulation for a longer period of time than would otherwise be possible. However, because the shock itself is treated as a discontinuity in the thermal plasma, the transition from thermal to non-thermal particles is not part of the model and needs to be prescribed based on information obtained from PIC simulations.

For this simulation we use a 2-D box-size with 180$\times$30\,$R_L$; $R_L$ being the Larmour radius defined by the velocity of the injected particles and the upstream magnetic field. This box is covered by a fixed grid with individual cells of 0.15\,$R_L$, which guarantees that the gyro motion of the individual particles is adequately resolved. 
Unlike the PIC simulations, we run this model in the reference frame of the shock, starting from an analytical solution for a standing shock with an upstream $\theta_B\,=\,30^o$ derived from the Rankine-Hugoniot conditions. The flow moves from the inner x-boundary toward the outer x-boundary. 
Once the simulation starts, we inject protons at the shock with a rate of two percent of the mass that is passing through the shock front, matching the results shown in PIC simulations as shown in [10]. We neglect non-thermal electrons. This allows us to scale the resolution of the grid cells to the proton-gyro radius without the need to resolve the much shorter electron gyro-radius. 

At injection, the particles are given a Lorentz factor that is three times that of the upstream flow, with the velocity isotropically distributed in the frame of the shock. 

We repeat the simulation with an upstream flow that has a Lorentz factor of 3.0 to test the code's ability to perform simulations for relativistic shocks. This also obliges us to reduce $\theta_B$ to ten degrees, because a $\Gamma\,=\,3$ shock would become superluminal at an angle of thirty degrees.

\begin{figure}
    \centering
    \includegraphics[width=0.49\linewidth]{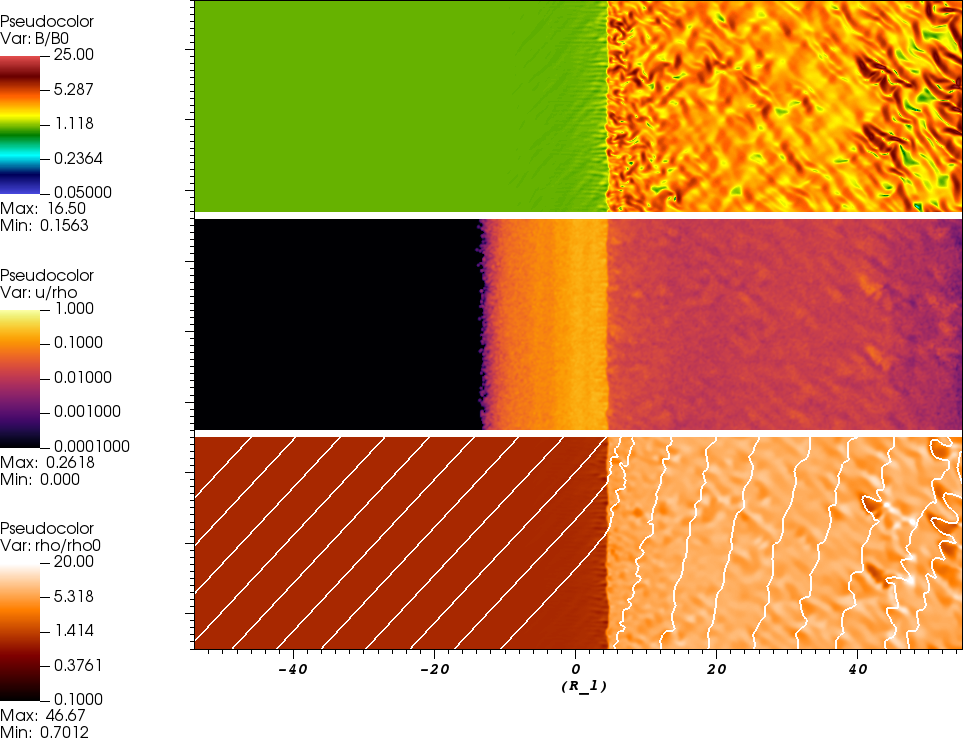}
    \includegraphics[width=0.49\linewidth]{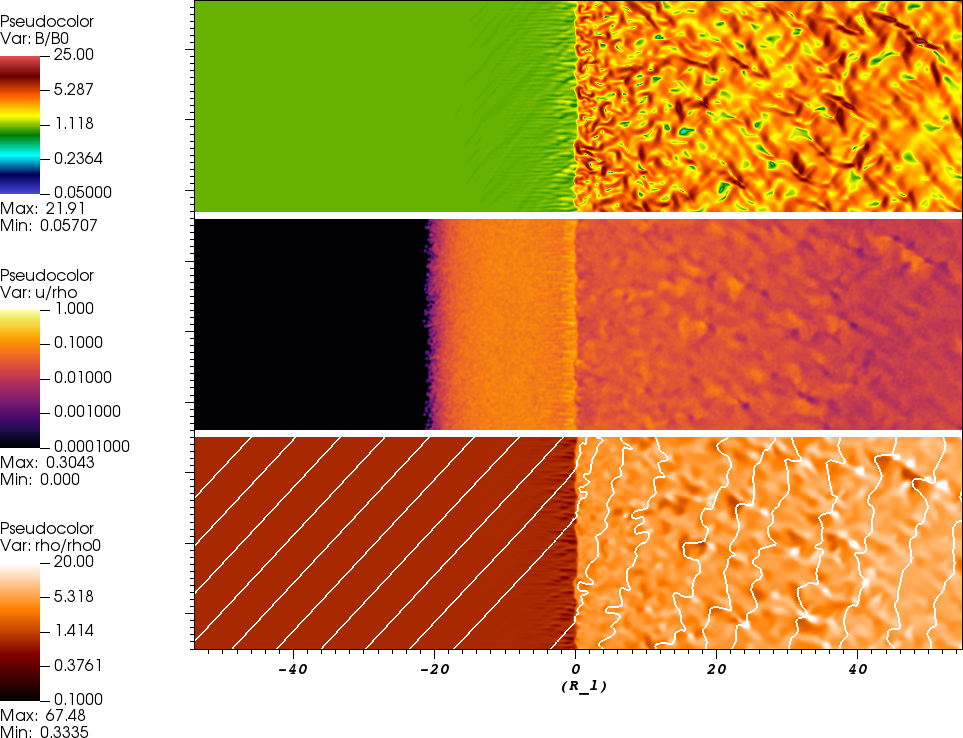} \\
    \includegraphics[width=0.49\linewidth]{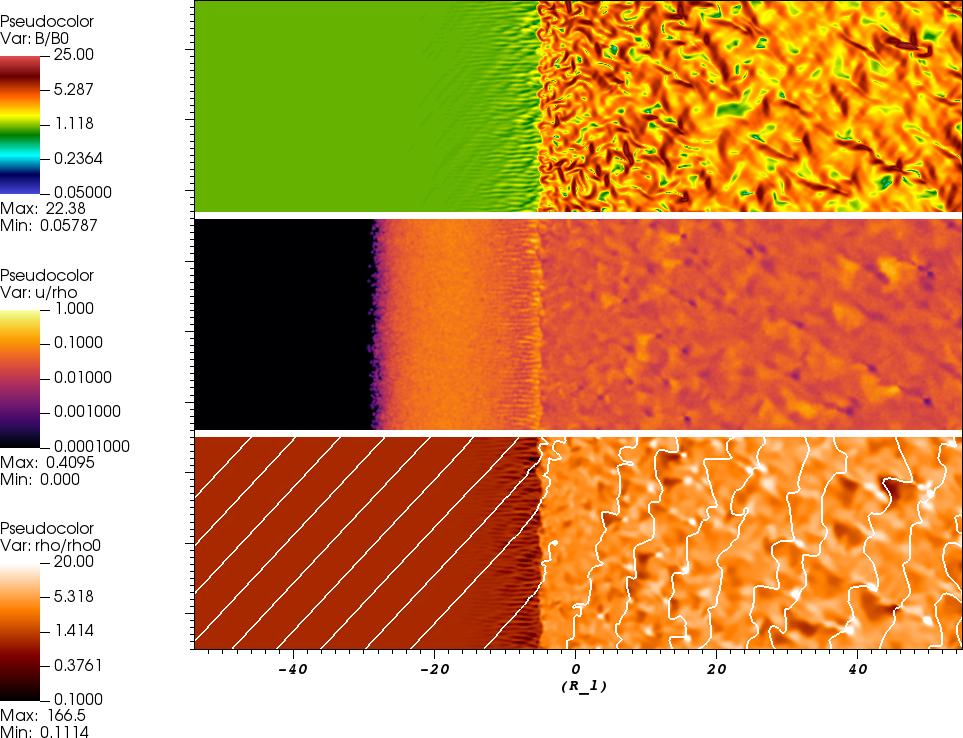}
    \includegraphics[width=0.49\linewidth]{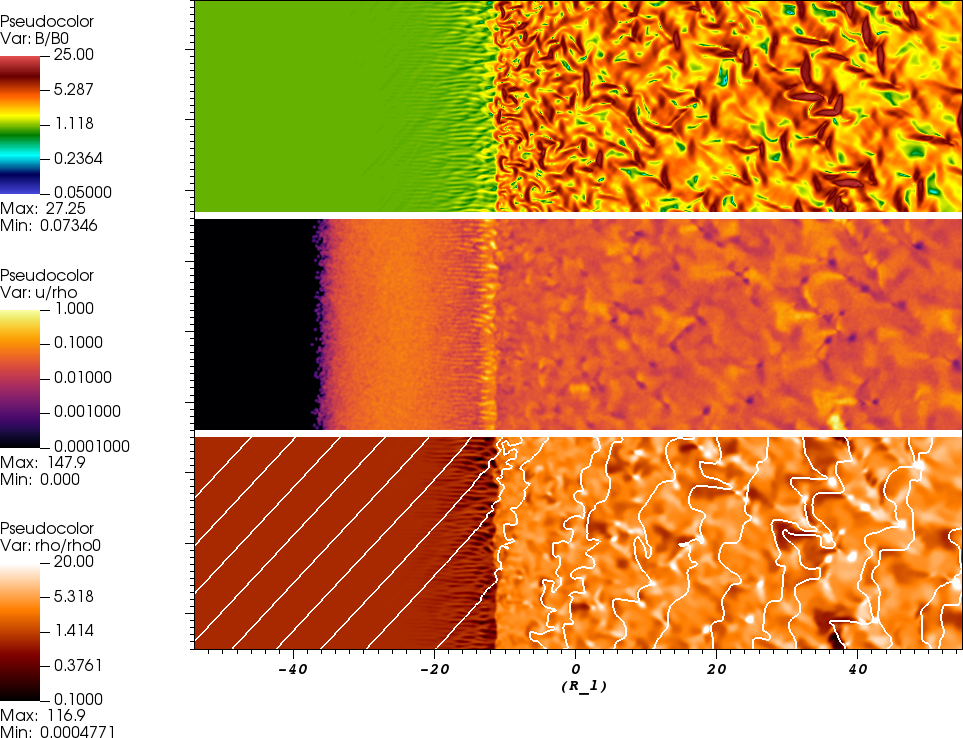} 
    \caption{Simulation results for a shock with $\Gamma\,=\,1.86$, $\theta_B\,=\,30^o$. From top to bottom, the figures show the magnetic field relative to the initial upstream field, the ratio of non-thermal to thermal particles, and the thermal gas density relative to the initial upstream density with the magnetic field lines for $t\,=\,200 \ R_L/c$ (upper left), $300\,R_L/c$ (upper right),$400\,R_L/c$ (lower left) and $500 \ R_L/c$ (lower right). }
    \label{fig:picmhdshock1}
\end{figure}

\section{Results}
\subsection{Simulation results for $\Gamma\,=\,1.86$}
Figure~\ref{fig:picmhdshock1} demonstrates the result of our simulation with $\Gamma\,=\,1.86$ and $\theta_B\,=\,30$. Here we show two snapshots in time at $t\,=\,200 \ R_L/c$ (upper left panel) through $500 \ R_L/c$ (lower right panel). Each panel shows the magnetic field relative to the initial upstream field (top), the non-thermal particle density relative to the thermal gas particle density (centre), and the thermal gas density relative to the initial upstream thermal gas density (bottom) as well as the magnetic field lines. 
In the early stage (upper left), the upstream magnetic field remains largely unperturbed. In any case, the particles have had no time to move far upstream. Downstream, the field lines show deviation from the initial condition, but the changes are still small. 
Over time, this changes. The upstream field shows signs of perturbation, which follows the expected form for a streaming instability though the deviations are relatively small, particularly when compared to non-relativistic results such as shown in [4,5,6]. In all of this, we should keep in mind that we do not include the effect of non-thermal electrons. As a result, we don't see electron-induced instabilities such as the Whistler and fire-hose instabilities.

\begin{figure}
    \centering
    \includegraphics[width=0.70\linewidth]{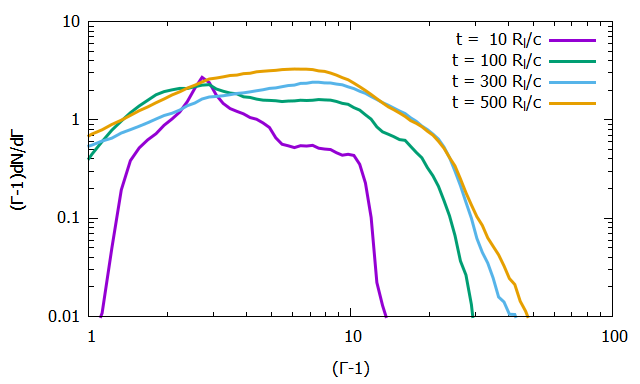}
    \caption{Time-evolution of the spectral energy distribution of non-thermal particles for a shock with $\Gamma\,=\,1.86$, $\theta_B\,=\,30^o$.}
    \label{fig:picmhdsed}
\end{figure}

The particle acceleration is demonstrated in fig.~\ref{fig:picmhdsed}, which shows the spectral energy distribution (SED) of the non-thermal particles as a function of time. 

To understand the evolution of the particle energies, we have to differentiate between two acceleration mechanisms. The first, DSA, was described previously. However, there is a second mechanism: Shock Drift Acceleration (SDA). Unlike DSA, which involves particles travelling large distances up- and down-stream before being reflected by local instabilities in the magnetic field, SDA is a localized phenomenon, which involves particles moving along the shock front, repeatedly transitioning between the upstream and downstream medium. It is effective in giving the particles an initial boost in energy. However, once a particle escapes from the vicinity of the shock, the process comes to an end, which limits the maximum energy that can be obtained. DSA, which involves a much larger volume of space, can continue as the particle gains energy as long as some instability catches the particle and reflects it toward the shock.

Initially, the SED shows only limited growth, ranging from non-relativistic to $\Gamma\,\simeq\,15$. In this stage, there is no DSA, but some particles are accelerated through SDA)]. SDA, which does not require large-scale instability in the upstream magnetic field, can occur because particles can repeatedly cross the shock as they circle the magnetic field lines. This process was shown in detail in [6]. While it would not be significant for a $\theta_B\,=\,30^o$ shock in a non-relativistic case, the Lorentz compression increases the effective angle of the upstream magnetic field in the shock rest frame, allowing for a more efficient SDA. 
Over time, the high-energy tail of the SED increases until it approaches $\Gamma\,=\,80$, demonstrating that the particles are being accelerated efficiently. We also see the initial stage of a power-law distribution, indicating the onset of DSA.

\subsection{Simulation results for $\Gamma\,=\,3$}
\begin{figure}
    \centering
    \includegraphics[width=0.49\linewidth]{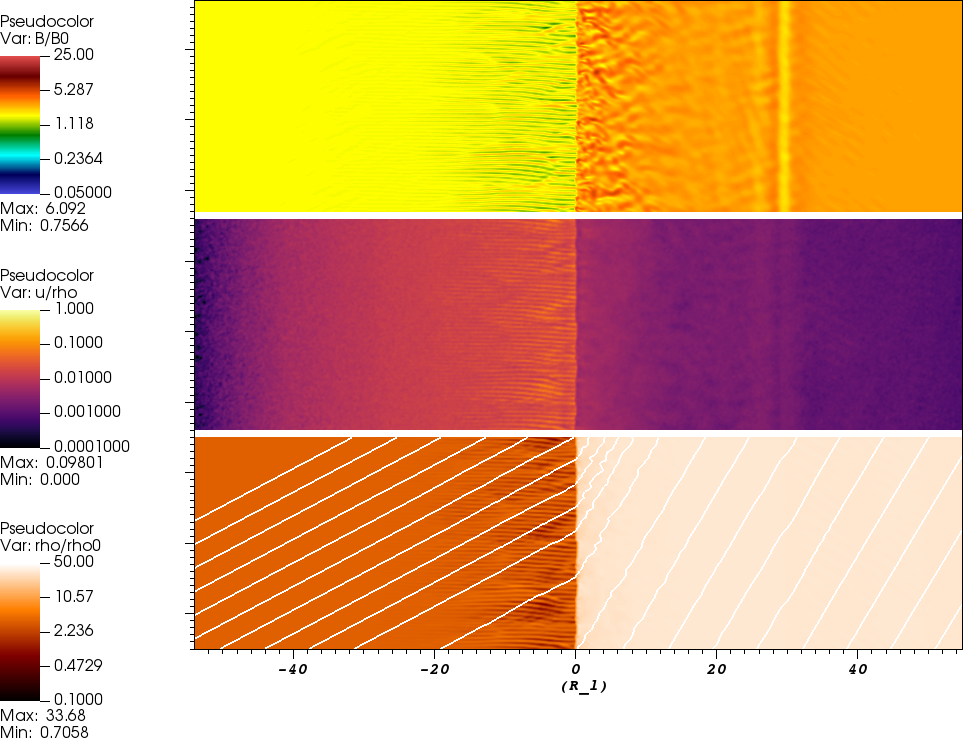}
    \includegraphics[width=0.49\linewidth]{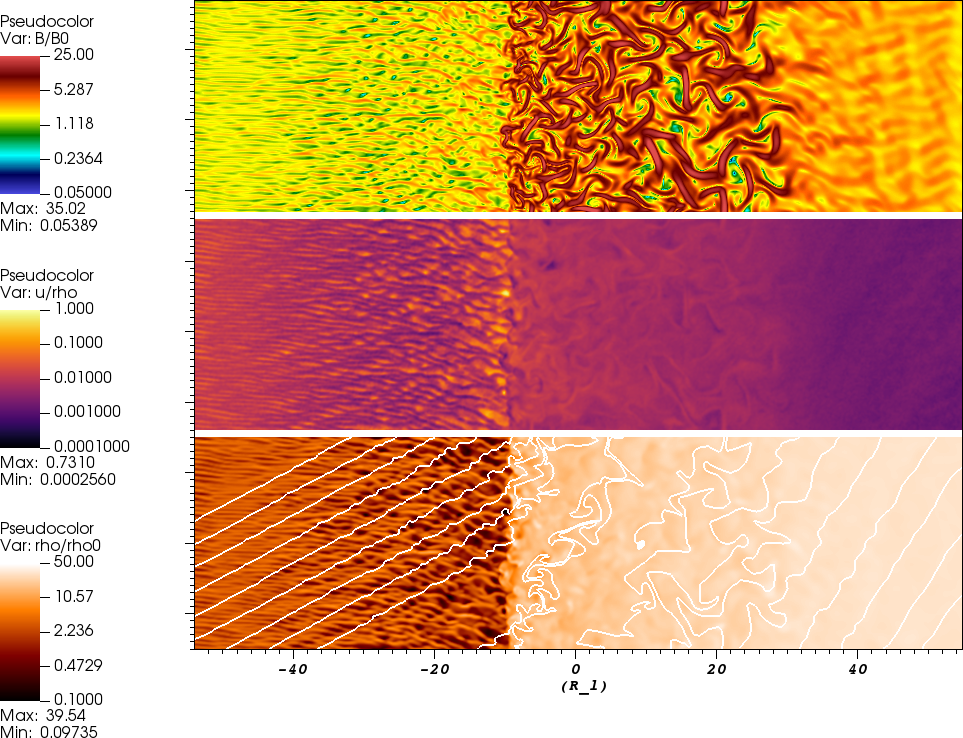} 
    \caption{Similar to Fig.~\ref{fig:picmhdshock1} but for a shock with $\Gamma\,=\,3$, $\theta_B\,=\,10^o$ for $t\,=\,100 \ R_L/c$ (left), $200\,R_L/c$ (right). }
    \label{fig:picmhdshock2}
\end{figure}

\begin{figure}
    \centering
    \includegraphics[width=0.70\linewidth]{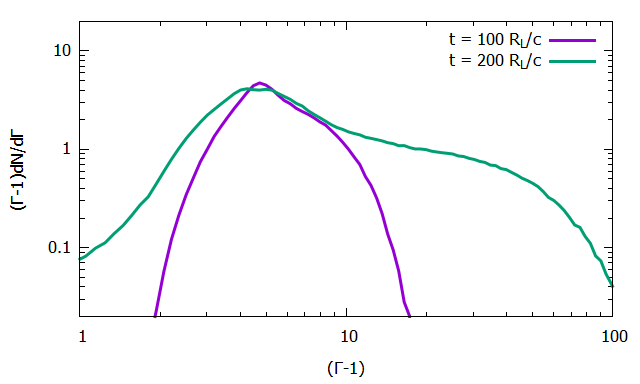}
    \caption{Time-evolution of the spectral energy distribution of non-thermal particles for a shock with $\Gamma\,=\,3$, $\theta_B\,=\,10^o$.}
    \label{fig:picmhdsed2}
\end{figure}

The results for our model with $\Gamma\,=\,3$ and $\theta_B\,=\,10^o$ are shown in  Fig.~\ref{fig:picmhdshock2}, which shows the same variables as Fig.~\ref{fig:picmhdshock1} at $t\,=\,100 R_L/c$ (left) and  $t\,=\,200 R_L/c$ (right).
Compared to the model with $\Gamma\,=\,1.86$, the instabilities develop very quickly, with significant upstream distortion of the magnetic field clearly visible at $t\,=\,200 R_L/c$. 
The SED for this simulation is shown in Fig.~\ref{fig:picmhdsed2}, which demonstrates that the acceleration is even more efficient than for the slower model, with particles exceeding $\Gamma\,=\,100$ within a short ($t\,=\,200 R_L/c$) timeframe.

\section{Conclusions}
By combining PIC and PIC-MHD simulations, we can create a model of an oblique, relativistic shock and investigate its ability to accelerate particles to the speed observed in cosmic rays. 
We find that for those shocks that are not superluminal, the injection rate is high. This allows the particles to create perturbations of the magnetic field in the upstream medium and trigger the non-resonant streaming instability. As a result, particles are accelerated, and we begin to observe the formation of a power-law SED. 
In fact, our results show that oblique, relativistic shocks can be highly efficient accelerators because of the combination of a high compression rate across the shock and the fact that the Lorentz contraction allows even mildly oblique shocks to cause SDA which can effectively combine with DSA.
From this, we can conclude that these shocks can contribute to the observed cosmic-ray spectrum.

In the future, we intend to extend our parameter space to higher relativistic shocks ($\Gamma\,\geq\, 10$), and, potentially, extend the model to 3-D.

\section*{Acknowledgments}
A.~B. was supported by the German Research Foundation (DFG) as part of the Excellence Strategy of the federal and state governments - EXC 2094 - 390783311. Computations were performed on the HPC system Raven at the Max Planck Computing and Data Facility.

\smallskip

\end{document}